\documentclass{LMCS}

\def\doi{8 (3:15) 2012}
\lmcsheading%
{\doi}
{1--19}
{}
{}
{Nov.~\phantom08, 2011}
{Sep.~13, 2012}
{}

\usepackage{enumerate}
\usepackage{hyperref}
\usepackage{tikz}

\theoremstyle{plain}

\def\IB{\mathbb{B}}
\def\IC{\mathbb{C}}

\def\IN{\mathbb{N}}
\def\IQ{\mathbb{Q}}
\def\IR{\mathbb{R}}

\newcommand{\rang}{\mbox{\rm range}}

\begin{document}

\title[Point-Separable Classes of Computable Curves]{Point-Separable Classes\\ of Simple Computable Planar Curves}

\author[X.~Zheng]{Xizhong Zheng\rsuper a}	
\address{{\lsuper a}Jiangsu University, Zhenjiang 212013, China, and
Arcadia University, Glenside, PA 19038, USA}	
\email{zhengx@arcadia.edu}
\thanks{{\lsuper a}The first author is supported by DFG  (446 CHV 113/266/0-1),  NSFC (10420130638) and NSFC 61070231}	

\author[R.~Rettinger]{Robert Rettinger}
\address{
FernUniversit\"at Hagen, 58084 Hagen, Germany\rsuper b}	
\email{{\lsuper b}robert.rettinger@FernUni-Hagen.de }
\thanks{{\lsuper b}The second author is supported by DFG  (446 CHV 113/266/0-1) and NSFC (10420130638)}


\keywords{Computable Curves, Point separable.}
\subjclass{F.1.3}
\titlecomment{Part of the results have been presented at CCA2009 and MFCS2009.}


\begin{abstract}
\noindent
In mathematics curves are typically defined as the images of continuous real functions ({\em parametrizations}) defined on a closed interval. They can also be defined as connected one-dimensional compact subsets of points. For simple curves of finite lengths, parametrizations can be further required to be injective or even length-normalized. All of these four approaches to curves are classically equivalent. In this paper we investigate four different versions of {\em computable curves} based on these four approaches. It turns out that they are all different, and hence, we get four different classes of computable curves. More interestingly, these four classes are even {\em point-separable} in the sense that the sets of points covered by computable curves of different versions are also different. However, if we consider only computable curves of computable lengths, then all four versions of computable curves become equivalent. This shows that the definition of computable curves is robust, at least for those of computable lengths. In addition, we  show that the class of computable curves of computable lengths is point-separable from the other four classes of computable curves.
\end{abstract}

\maketitle

\section{Introduction}\label{sec-intro}

A curve is a mathematical model which describes the ``path (or locus) of a continuously moving point". Therefore, a planar curve is defined in mathematics as the image of a continuous function $f: [0, 1] \to \IR^2$. Surprisingly, under this definition, a curve can be so complicated that it fills even a square (cf. \cite{Peano1890,EHMoo1900}). In fact, as shown independently by Hahn and Mazurkiewicz in about 1913, a point set is a curve if and only if it is a locally connected continuum (we ignore the mathematical details in this paper which are not related to our discussion). However, if we are interested only in the curves which do not cross themselves (i.e., {\em simple}) and have finite length (i.e., {\em rectifiable}), then the curves defined in these ways coincide with our intuition about ``curves" and they do have the ``two-sidedness'' and ``thinness" (cf \cite{Why42}). For rectifiable simple curves, the parametrizations can be required to be injective or length-normalized while the induced class of curves remains the same. Therefore, a rectifiable simple curve can be defined as any of the following: a point set of some special topological properties, the image of a continuous function, the image of an injective continuous function, or the image of a continuous function which is length-normalized.

If a point-movement is ``algorithmically determined'', then its path (the curve) should be considered ``computable''. More precisely, the notion of computable curves can be defined by the effectivization of the classical definition of curves. This naturally raises the question whether the effectivizations of these four definitions of curves mentioned above lead to the same notion of ``computable curves''? Our answer is no, even in a very strong sense. Before we can explain our answer in a more precise way, let us recall first the basic idea of how to define computability of continuous objects in general.

In computable analysis, computability over various continuous structures is typically defined by the Turing-machine-based bit model (see \cite{Ko91,Wei00,BC06}). In order to input a real number $x$ to such a Turing machine, it must be represented by effectively convergent sequences of rational numbers (the {\em names} of $x$). Here, a sequence $(x_n)$ ``converges effectively" means that $|x_n-x_{n+1}| \le 2^{-n}$ for all $n$. A real number $x$ is computable if it has a computable name, i.e., there is a computable sequence of rational numbers which converges to $x$ effectively. Furthermore, a real function $f$ is computable if there is a Turing machine which computes $f$ in the sense that, after inputing any name of a real number $x$ in the domain of $f$, the machine outputs a name of $f(x)$. By the same principle, computability of other mathematical objects can be defined by introducing proper ``naming systems". For example, the computability of subsets of the Euclidean space \cite{BW99}, of semi-continuous functions \cite{WZ97}, of functional spaces \cite{ZW03} are all defined in this way. This approach is also called the ``effectivization" of classical mathematical definitions.

The same approach can be applied to curves as well. In this paper, we only consider plane curves. Curves in higher dimensions can be discussed in essentially the same way. Furthermore we will restrict ourselfes to rectifiable curves unless otherwise said, where a curve is {\em rectifiable} if it has a finite length.
As mentioned above, a curve can be defined as a connected and one-dimensional compact subset. Based on this approach, we can define the computable curves by means of the computability of compact subsets of Euclidean space (\cite{BW99}).  Physically, a curve records the trace of a particle motion. If the particle moves according to some algorithmically definable laws, its trace should be regarded as computable. In mathematical terms, a curve is the range of a continuous function defined on a closed interval and this function is called a parametrization of the curve. Thus, it is also natural to call a curve computable if it has a computable parametrization (see e.g.,  \cite{GLM06,GLM11}).

However, the parametrization of a curve may have various extra properties, if the curve is simple. Here a curve is called {\em simple} if it does not intersect itself, or if it has an injective parametrization. Of course, the parametrization $f$ of a simple curve $C$ is not necessarily injective. If $f$ is not injective, then $f$ retraces some parts of the curve $C$. If a curve $C$ is simple, then it has even an arc-length normalized parametrization. Here, a parametrization  $f:[0,1] \to C$ is arc-length normalized roughly means that the function $f$ models a particle movement along the curve $C$ with a constant speed.

In this paper, four versions of computable curves are introduced by effectivizing the above four mathematical approaches to curves. We will see that these four versions of computable curves are all different. The difference of the curve classes defined by simple computable parametrizations and computable injective parametrizations was already shown by Gu, Lutz and Mayordomo in a recent paper \cite{GLM11}. However, in this paper we will distinguish these four versions of computable curves in a much stronger sense. Namely, the sets of points covered by the four classes of computable curves are different. In other words, different versions of computable curves can be separated by the points they cover, or they are ``point-separable" (see definition in Section \ref{sec-point-separable}).

Interestingly, the computability of the curve length plays an important role for the computability of curves. If we look only at curves of computable lengths, then the four effectivizations mentioned above are indeed equivalent. This means that the definition of computable curves is robust, at least, for curves of computable lengths.  On the other hand, Gu, Lutz and Mayordomo  constructed  in \cite{GLM11} a computable curve of non-computable length such that none of its computable parametrizations can be injective, although the curve does not intersect itself. As an open question, they asked whether {\em there exists a point which lies on a computable curve of finite length, but not on any computable curve of computable length}, i.e., if the class of computable curves of computable lengths is point-separable from the class of computable curves of finite lengths. A positive answer will be given in this paper.

Our paper is organized as follows. In Section \ref{sec-comp-curve} we will briefly recall some basic notions related to curves, give the precise definition of computable curves and then show some basic properties of computable curves. In Section \ref{sec-point-separable}, we discuss some basic facts of point-separable classes and show a technical lemma which will be used in the proof of the main theorems. Section \ref{sec-n-comp-curve} investigates the class of length-normalized computable curves and shows a significant difference between this class and the class of computable curves of computable lengths. Then it is shown that  these two  classes are point-separable. In the last Section \ref{sec-main} we prove that the four classes of computable curves mentioned above are all point-separable.


\section{Computable Curves}\label{sec-comp-curve}

In mathematics, a {\em plane curve} is defined as a subset $C \subseteq \IR^2$ which is the range of a continuous function $f:[0;1]\to \IR^2$, i.e., $C = {\rm range}(f)$. This continuous function $f$ is then called a {\em parametrization} of $C$. Here, we use, w.l.o.g., the unit interval $[0,1]$ instead of more general closed intervals of the form $[a,b]$. Obviously, any curve has infinitely many parametrizations. Geometrically, a curve records the path of a particle movement in the plane. If the particle never visits one position more than once, in other words, if the curve does not intersect itself (or it has an injective parametrization $f:[0;1] \to \IR^2$), then the curve is called {\em simple}. The simple curves defined in this way are also called {\em open}, or {\em Jordan arcs}. If a curve $C$ has a parametrization $f$ which is injective on the interval $[0;1)$ and fulfills the condition that $f(0)=f(1)$, then the curve $C$ is traditionally also called {\em simple}, but it is {\em closed}, or a {\em Jordan curve}. Equivalently, a Jordan curve is the continuous image of the unit circle. In this paper we look only at the open simple curves. But all results are true for closed simple curves as well.

For open simple curves, their lengths can be defined by means of the lengths of polygons which approximate the curves according to Jordan \cite{Jor1882}. More precisely, Let $C$ be a simple curve and let $f:[0;1] \to \IR^2$ be an injective continuous parametrization of $C$. Then the {\em length} $L$ of the curve $C$ is defined by
\begin{eqnarray}\label{def-arc-length}
L := \sup \sum_{i=0}^{n-1} |f(a_i) -f(a_{i+1})|
\end{eqnarray}
where $|f(a_i) -f(a_{i+1})|$ is the length of the straight line connecting the points $f(a_i)$ and $f(a_{i+1})$, and the supremum is taken over all possible partitions $0=a_0<a_1<...<a_n=1$ of the unit interval $[0,1]$. The length of a curve $C$ is denoted by $l(C)$. Notice that we actually defined the length $l(f)$ of the function $f:[0;1]\rightarrow \IR^2$. The length of a simple curve is then the length of an injective parametrization of that curve. It is well known that the length of a simple curve is independent from its (injective) representations.  A curve of finite length is traditionally called {\em rectifiable}. Not every curve, even a simple curve,  has finite length. As already mentioned above, we focus mainly on simple rectifiable curves; unless otherwise stated a curve is always meant to be simple and rectifiable in this paper.

If $C$ is a simple rectifiable curve of the length $l$, then there exists a bijective continuous function $g:[0, l] \to C$ such that the arc $g([0, s])$ has exactly the length $s$. That is, the arc-length $s$ is used as the argument of the function $g$. Let $f(t) := g(l\cdot t)$. Then the function $f:[0,1]\to C$ is a parametrization such that the curve segment $f([0, t])$ has the length $t\cdot l(C)$ for all $t\in[0,1]$. We call the parametrization $f$ of this property {\em length-normalized} or simply {\em normalized}.  Thus, a simple rectifiable curve can have three different kinds of parametrizations---continuous, injective continuous and normalized. In addition, a curve can also be defined as a connected, one-dimensional, compact point set. By effectivizing  all these approaches to curves, we can introduce four totally different versions of computable curves.

Remember that a real function $f:[0; 1] \to \IR$ is computable if there is a Turing machine $M$ which transfers any name of $x\in[0,1]$ to a name of $f(x)$.  Equivalently, $f$ is computable iff there is a computable sequence $(p_n)_{n\in\IN}$ of computable rational polygon functions which converges uniformly and effectively to $f$ (see \cite{PR89}). Naturally, a function $f:[0;1] \to \IR^n$ is computable if all of its component functions are computable, or equivalently, if there is a Turing machine $M$ which transfers any name of $x\in [0,1]$ into a tuple $(\alpha_1, \cdots, \alpha_n)$ of names of $f_1(x), \cdots, f_n(x)$ respectively, where $f(x) = (f_1(x), \cdots, f_n(x))$. In this case, we simply say that $M$ computes the function $f$. Remember also that any computable function must be continuous.

In this paper, an $\varepsilon$-neighborhood $V_\varepsilon(z)$ of a point $z=(a,b)$ with Cartesian coordinates $(a,b)$ is the rectangle bounded by the lines $x= a \pm \varepsilon$ and $y = b\pm \varepsilon$. A neighborhood $V_\varepsilon(z)$ is called rational if $z$ is a rational point and $\varepsilon$ is a rational number. For a set $A \subseteq \IR^2$, the $\varepsilon$-neighborhood of $A$ is defined by $V_\varepsilon (A) := \bigcup_{z \in A} V_\varepsilon(z)$. Then for any two point sets $A, B$, their Hausdorff distance is defined by $d_H(A, B) = \inf \{ \varepsilon: A \subseteq V_\varepsilon(B) \ \&\ B \subseteq V_\varepsilon(A)\}$. Notice that, we  always have $d_H(V_\varepsilon(z), z) \le \sqrt{2}\varepsilon$.

Now we can define the different versions of computable curves as follows.

\begin{defi}\label{Def-comp-curve}
Let $C \subseteq \IR^2$ be a simple, not necessarily rectifiable, planar curve.
\begin{enumerate}[(1)]
 \item $C$ is called {\em $K$-computable} if there is a computable sequence $(Q_n)$ of finite sets of rational neighborhoods such that
\begin{eqnarray}\label{cond-K-comp}
C \subseteq \bigcup Q_n &\mbox{and} & d_H\left(\bigcup Q_n,\,\,C\right)<2^{-n}
\end{eqnarray}
for all $n\in \IN$, where $d_H$ denotes the Hausdorff distance.

\item $C$ is called {\em $R$-computable} if there is a computable function $f:[0;1]\rightarrow \IR^2$ such that $\rang(f) =C$.

\item $C$ is called {\em $M$-computable} if there is an injective computable function $f:[0;1]\rightarrow \IR^2$ such that $\rang(f) =C$.

\item $C$ is called {\em $N$-computable} if $C$ has a computable parametrization $f:[0;1]\rightarrow \IR^2$ such that the length of the curve segment $f([0, t])$ is equal to $t\cdot l(C)$ for all $t\in [0,1]$.
\end{enumerate}
\end{defi}

\noindent In item (1) of the definition, the finite sets $Q_n$ of rational neighborhoods are also called compact covers of the curve $C$. The union $\bigcup Q_n$ means the union of all neighborhoods in $Q_n$, not the union $\bigcup_{n\in\IN} Q_n$.   The second part of condition (\ref{cond-K-comp}) means that the maximum distance from $C$ to the boundary of the compact cover $Q_n$ is bounded by $2^{-n}$. W.l.o.g., we can even require that the sequence $(Q_n)$ is decreasing in the sense that $\bigcup Q_{n+1} \subseteq \bigcup Q_n$ for all $n$.  The letter $K$ of the $K$-computability comes from the German word $K$ompakt (compact) due to the compact coverings.

In item (2), the letter $R$ stands for  $R$etracable because the parametrization $f$ of a $R$-computable curve $C$ can retrace the curve $C$. Namely, there might be some disjoint subintervals $I_1, I_2 \subset [0,1]$ such that $f(I_1) = f(I_2)$. In this case, $f$ traces the segment $f(I_1)$ of $C$ more than once, or, we say that $f$ is retraceable.

If the parametrization of a curve $C$ is injective, then $C$ records the movement of a particle with a monotone direction. The letter $M$ in $M$-computability stands for $M$onotonically  directed movement or $M$onotone paramatrization. Notice that, if we consider also closed simple curves, then the monotonicity has to exclude the endpoints of the unit interval.

Finally, if a parametrization $f:[0,1] \to \IR^2$ satisfies the condition that the length of the curve segment $f([0, t])$ is proportional to $t$, i.e., $l(f([0, t])) = t\cdot l(C)$ for all $t\in [0,1]$,  then it is  normalized. Thus, $N$-computability stands for $N$ormalized parametrization.

By definition \ref{Def-comp-curve}, any $N$-computable curve must be rectifiable. However, it is known that $M$-computable curves can have infinite lengths
(see e.g \cite{Ko98}). We will give a simple proof of this fact below by constructing a  Koch curve, which is well known to be $M$-computable (see e.g. \cite{Kam96}). The main reason for re-proving the following result is to introduce basic curve construction techniques which will be used throughout the more involved proofs in the next sections.

\begin{thm}\label{thm-inf-length-M-curve}
There is an $M$-computable curve $C$ which has infinite length.
\end{thm}

\proof
We will construct a computable sequence $(p_n)$ of rational polygons inductively and finally let $C$ be the limiting curve of this sequence. Here, a rational polygon is simply a finite sequence $[q_0,...,q_r]$ of rational points $q_i\in\IQ^2$ and its (not necessarily simple) curve is the union of all line segments connecting these points in their given order. We use the term polygon to mean both the point sequence and the corresponding curve. In the following we will construct a new polygon $p_{n+1}$ from $p_n= [q_0,...,q_r]$ by adding new points to the sequence $[q_0,...,q_r]$ without deleting the original points or changing their relative order.

Given a polygon $p=[q_0,...,q_r]$ we can define straightforwardly its length-normalized parametrization $\hat {p}: [0,1] \to \IR^2$ by

\begin{eqnarray*}
\hat{p}(t) = q_i + \frac{t-t_i}{t_{i+1} -t_i}(q_{i+1} -q_{i}) \qquad \mbox{ for } t \in [t_i ,\ t_{i+1}]
\end{eqnarray*}
where $t_0 =0$ and
\begin{eqnarray*}
t_i = \frac{\sum_{j=0}^{i-1} \left|q_j- q_{j+1}\right|}{\sum_{j=0}^{r-1} \left| q_j - q_{j+1}\right|} \qquad \mbox{ for all } 0 <i \le r.
\end{eqnarray*}

Back to the proof of our theorem, we construct the sequence $(p_n)$ of polygons as follows:
Let $p_0 =[(0,0),\ (1,0)]$. Then we define $p_1:=[(0,0), \ (1/4, 1/4), \ (1/2, 0), \ (3/4, -1/4), \ (1, 0)]$ by adding three new points $(1/4, 1/4), \ (1/2, 0), \ (3/4, -1/4)$ to $p_0$. Thus $p_1$ consists of four line segments of length $\sqrt{2}/4$ and it has a total length $\sqrt{2}$, i.e., $l(p_1) = \sqrt{2}$.  Apparently we have $d_H(p_0, p_1)= 1/4 $ and $|\hat{p}_0(t) -\hat{p}_1(t)| \le 1/4$ for all $t\in [0,1]$.

A similar procedure can be applied to each of the four segments of $p_1$ to construct a polygon $p_2$ consisting of 16 segments of the length $(\sqrt{2}/4)^2$ and hence $l(p_2) = (\sqrt{2})^2$. In addition, we have  $|\hat{p}_1(t)-\hat{p}_2(t)| \le {\sqrt{2}}/{4^2}$ for all $t\in [0,1]$. Continuing this process inductively, we can construct a computable sequence $(p_n)$ of rational polygons\footnote{It is possible that some polygons contains non-rational points by this construction. But these points can only be algebraic. In this case these irrational points can be replaced by some close enough rational points to guarantee that the result holds as well. For the simplicity, we skip the details here. } such that
\begin{equation} \label{RoEq1}
l(p_n) = \left(\sqrt{2}\,\right)^n \quad \mbox{ and } \quad \left|\hat{p}_n(t)- \hat{p}_{n+1}(t)\right| \le \frac{(\sqrt{2})^n}{ 4^{(n+1)}} \le 2^{-n}
\end{equation}
for all $n\in\IN$ and $t\in[0,1]$.

The second part of condition (\ref{RoEq1}) implies that the limit $f(t) =\lim_{n\to\infty} \hat{p}_n(t)$ exists and it is computable, and hence a continuous function which should be a parameterization of the limiting curve  $C := \lim p_n$. By definition of the curve length, we have $l(C) \ge l(p_n) = (\sqrt{2})^n$ for all $n$ because $p_n =[\hat{p}_n(0),\cdots, \hat{p}_n(i\cdot 2^{-2n}), \cdots, \hat{p}_n(2^{2n}2^{-2n})]$ and  $f(i\cdot 2^{-2n}) = \hat{p}_n(i\cdot 2^{-2n})$ for all $0\le i \le 2^{2n}$. Therefore $C$ has an infinite length.

It remains only to be shown that $f$ is also injective. This follows immediately from the fact that $|\hat{p}_n(t_1)-\hat{p}_n(t_2)|\geq |t_1-t_2|/3$ which can be proved by induction on $n$. By the uniform convergence of the sequence $(\hat{p}_n)$, we conclude that $|f(t_1) -f(t_2)| \ge |t_1 -t_2|/3 $, that is, $f$ is an injective parameterization of $C$ and hence $C$ is an $M$-computable curve.
\qed

Although a computable curve may have infinite length, computable rectifiable curves seem more interesting and more important. As mentioned above we will focus on computable curves of finite length in this paper and we denote by $\IC_K,\IC_R, \IC_M$ and $ \IC_N$ the classes of all $K$-, $R$-, $M$- and $N$-computable rectifiable simple curves, respectively. By definition, it is straightforward that we have the following relationship between these four versions of computable curves.

\begin{thm}\label{Thm-comp-curve-subset}
$\IC_N \subseteq \IC_M \subseteq \IC_R \subseteq \IC_K$.
\end{thm}
We will see that all four versions of  computable curves are different and hence all the subset relations above are proper.

From (\ref{def-arc-length}) it is straightforward that the length of a rectifiable $M$-computable curve is left computable (see also Theorem of \cite{MZ2008}), where a real number $x$ is left computable, or computably enumerable (c.e. for short), if there is an increasing computable sequence $(x_n)$ of rational numbers which converges to $x$. In \cite{GLM11}, Gu, Lutz and Mayordomo have shown that any rectifiable $R$-computable curve also has a left computable length. This can be strengthened further to the $K$-computable curves as follows.

\begin{thm}\label{thm-lc-length-K}
Any rectifiable $K$-computable curve has left computable length.
\end{thm}
\proof If $C$ is a rectifiable $K$-computable curve, then there is a decreasing computable sequence $(Q_n)$ of rational compact covers of $C$ such that $d_H\left(\bigcup Q_n,\,\,C\right)<2^{-(n+1)}$ where $Q_n$ is a finite set of rational neighborhoods for all $n$. Furthermore let $\overline{Q}_n$ be the corresponding finite set of the closed coverings where each open neighborhood of $Q_n$ is replaced by its closure. Thus $\overline{U}:= \bigcup \overline{Q}_n$ (the union of all sets in $\overline{Q}_n$) is a closed rational polygon area built of rational neighborhoods (squares). Since this area contains at least one curve $C$ such that $d_H(C, \overline{U}) = d_H(C, \bigcup Q_n) \le 2^{-(n+1)}$, we can find, for each $n$, a simple polygon $p_n$ of shortest length in this area such that $d_H(p_n, \overline{U}) \le 2^{-n}$.
Let $l_n$ be the length of $p_n$ and let $l'_n$ be some rational approximation of $l_n$ with $l_n-2^{-n}\leq l'_n\leq l_n$. Thus $(l'_n)$ is a computable sequence of rational numbers. Obviously we have $l'_n \le l(C)$.

In the following we will prove that the length $l_n$ of $p_n$ (and hence also $l'_n$) will be  arbitrarily close to $l(C)$. Therefore, $l(C) = \lim_{n\to\infty} \max\{l'_i: i\le n\}$ is left computable.

Let $f$ be an injective (not necessarily computable) parametrization of $C$. By definition of curve length, for any $\varepsilon >0$, there exists a partition $0=t_0<t_1<\cdots < t_m=1$ such that $\sum_{i=0}^{m-1} \left| f(t_i)-f(t_{i+1})\right| \ge l(C) -\varepsilon/2$. Let $q =[f(t_0), f(t_1), \cdots, f(t_m)]$ be the corresponding polygon. Then we have $l(q) \ge l(C)-\varepsilon/2$.

We try now to compare the lengths of the polygons $q$ and $p_n$ for large indices $n$. Let
\begin{eqnarray*}
\delta_i := \max\left\{d_H(f(t_i),\, f([0, t_{i-1}])),\ d_H(f(t_i),\, f([t_{i+1}, 1]))\right\}
\end{eqnarray*}
for $i \le m$. Let $t_{-1}=-1$ and $t_{m+1} =2$ for technical reasons. Consider the $\delta$-neighborhoods $U_i= V_\delta(f(t_i))$  of $f(t_i)$, where $\delta := \min\{\delta_0/4, \delta_1/4, \cdots, \delta_{m}/4, \varepsilon/(4\sqrt{2}m)\}$. Notice that, the $\delta$ is small enough such that $U_i \cap U_j = \emptyset$  if $i \neq j$.  Choose an index $n$ large enough such that $2^{-n} \le \delta/2$ and consider the rational compact cover $Q_n$ with $d_H(C, \overline{U})\le 2^{-n}$ where $\overline{U} =\bigcup \overline{Q}_n$. Remember that, $p_n$ is a rational polygon in $\overline{U}$ of the shortest length such that $d_H(p_n, \overline{U}) \le 2^{-n}$. This, together with $d_H(C, \overline{U})\le 2^{-n}$, implies that $d_H(C, p_n) \le 2^{-n+1} \le \delta$. In particular, we have $d_H(f(t_i), p_n)\le \delta$ for all $i\le m$ which implies that $\overline{U}_i \cap p_n \neq \emptyset$. That is, there are $s_i\in [0,1]$ such that $\hat{p}_n(s_i) \in \overline{U}_i$ for all $i\le m$, where $\hat{p}_n$ is a length-normalized parametrization of $p_n$. Notice that $U_i$ and $U_{i+1}$ are disjoint neighborhoods but they are connected by a  subarea of $\overline{U}$ containing the curve segment $f[t_i, t_{i+1}]$ of $C$. For the neighborhoods $U_i$ and $U_{i+2}$, they are also disjoint, and the only possible path in $\overline{U}$ which connects them must pass through the neighborhood $U_{i+1}$. All shortcut between $U_i$ and $U_{i+2}$ without passing through $U_{i+1}$ will have a Hausdorff distance greater than $\delta$. This is generally true for any non-neighboured  $U_i$ and $U_j$ (i.e., $|i-j| \ge 2$). Therefore, the polygon $p_n$ can connect points $\hat{p}_n(s_i)$ only in the order $\hat{p}_n(s_0), \hat{p}_n(s_1), \cdots, \hat{p}_n(s_m)$ (or the reverse one). W.l.o.g., we can assume that $s_0<s_1<\cdots <s_m$. For any $i$, the polygon $q$ connects two points $f(t_i)$ and $f(t_{i+1})$ by a straight line, while the polygon $p_n$ may connect the points $\hat{p}_n(s_i)$ and $\hat{p}_n(s_{i+1})$ by several linear segments. Therefore, we have $l(f[t_i, t_{i+1}])\le l(\hat{p}_n[s_i, s_{i+1}]) +2\sqrt{2}\delta$ because $\hat{p}_n(s_i)$ is in the $\delta$-neighborhood $\overline{U}_i$ of $f(t_i)$,  and hence $d_H(f(t_i), \, \hat{p}_n(s_i)) \le \sqrt{2}\delta$. This implies that $l(q) \le l(p_n) + 2\sqrt{2}m\delta\le l(p_n)+\varepsilon/2$. Thefore, we can conclude that $l(C) \le l(q) +\varepsilon/2 \le l(p_n) +\varepsilon$. Since $\varepsilon$ is arbitrary, we have $\lim l_n =l(C)$
\qed

The construction in the proof of Theorem \ref{thm-lc-length-K} implies immediately an equivalent characterization of $K$-computable curves as follows.

\begin{cor}\label{cor-K-equivalence}
A rectifiable curve $C$ is $K$-computable if and only if there is a computable sequence $(p_n)$ of rational polygons which converges to $C$ effectively in the sense that $d_H(p_n, p_{n+1}) \le 2^{-n}$.
\end{cor}
\proof
If $C$ is a $K$-computable curve, then there is a computable sequence $(Q_n)$ of rational compact covers of $C$. By a construction given in the proof of Theorem \ref{thm-lc-length-K}, there is a computable sequence $(p_n)$ of rational polygons which converges to $C$ effectively.

On the other hand, if $(p_n)$ is a computable sequence of rational polygons which converges to $C$ effectively, then we have $d_H(C, p_{n+1}) \le 2^{-n}$. Construct a rational compact cover $Q_n$ of $p_{n+1}$ such that $d_H(p_{n+1}, \bigcup Q_n) \le 2^{-n}$. Then $Q_n$ is also a rational compact cover of $C$ such that $d_H(C ,\bigcup Q_n) \le 2^{-n+1}$. That is, $C$ is $K$-computable.
\qed

By Theorems \ref{Thm-comp-curve-subset} and \ref{thm-lc-length-K}, any rectifiable  $R$-, $M$- and $N$-computable curve has also a left computable length. Ko \cite{Ko95a} constructed a ``monster curve" which is $M$-computable (even in polynomial time) with a non-computable length. This implies that the length of a $K$-computable curve is not necessarily computable. Our next theorem shows that the computability of the curve-length plays a very important role in the study of computable curves.

\begin{thm}\label{thm-K+compLengh}
If $C$ is a $K$-computable curve  with a computable length, then $C$ must be $N$-computable.
\end{thm}
\proof
Suppose that $C$ is a $K$-computable curve whose length $l =l(C)$ is a computable real number. Then there is a decreasing computable sequence $(Q_n)$ of rational compact covers of $C$ such that $C \subseteq \bigcup Q_n$ and $d_H(C,\, \bigcup Q_n) \le 2^{-(n+1)}$. There is also an increasing computable sequence $(r_n)$ of rational numbers converging to $l$ effectively in the sense that $(r_{n+1}-r_n) \le 2^{-(n+1)}$. By the proof of Theorem \ref{thm-lc-length-K}, there exists a computable sequence $(p_n)$ of rational polygons such that $d_H(C, p_n) \le 2^{-(n+1)}$ for all $n$ and $\lim_{n\to\infty} l_n = l$ where $l_n := l(p_n)$. Notice that, because $(Q_n)$ is decreasing, the sequence $(l_n)$ is increasing. Furthermore, we have also that $d_H(p_n, p_{n+1}) \le 2^{-n}$ for all $n$.

For each $n\in\IN$, we can find a sufficiently large index $s_n$ such that  $|l_{s_n} -r_{s_n}| \le 2^{-(n+2)}$. Such an index $s_n$ exists because both sequences $(l_s)$ and $(r_s)$ converge to the same limit $l(C)$. Actually we can choose the sequence $(s_n)$ to be strictly increasing and $n < s_n$. Thus we have $|l_{{s_n}}- l_{s_{n+1}}| \le 2^{-(n+1)}$.  Since $p_{s_n}$ is a rational polygon, there is a computable function $f_n:[0,1]\to \IR^2$ such that $f_n$ is a length-normalized parametrization of $p_{s_n}$. Because of the conditions $d_H(p_{s_n}, p_{s_{n+1}}) \le 2^{-(n+1)}$ and $|l_{s_n} -l_{s_{n+1}}| \le 2^{-(n+1)}$, we can choose the computable sequence of functions $(f_n)$ such that $|f_n(t) -f_{n+1}(t)| \le 2^{-(n+1)}$ for all $t \in [0,1]$. In other words, the sequence $(f_n)$ converges effectively and hence its limit $f$ is also a computable function which is a length normalized parametrization of $C$. Therefore, the curve $C$ is  $N$-computable.
\qed

The following corollary follows immediately from Theorem \ref{thm-K+compLengh}.

\begin{cor}\label{cor-equi-compLength}
If $C$ is a rectifiable simple curve of computable length, then $K$-, $R$-, $M$-, and $N$-computability of $C$ are equivalent.
\end{cor}

Thus, if we consider only curves of computable length, then it is not necessary to distinguish between $K$-, $R$-, $M$- and $N$-computability of curves. That is, the notion of ``computable curves" is quite robust at least for simple curves of computable lengths.
Therefore, we can denote simply by $\IC_C$ the class of computable curves of computable lengths in any of these versions. Later on, we will call a curve {\em computable} (without mentioning the prefixes $K$, $R$, $M$ or $N$) if it is an element of $\IC_C$.

Now let $C$ be an $M$-computable rectifiable curve which is not $N$-computable (such curve exists by Theorem \ref{thm-M-N-sep}). This curve $C$ is of course also $K$-computable (Theorem \ref{Thm-comp-curve-subset}). By Theorem \ref{thm-K+compLengh}, $C$ does not have computable length. Therefore, there exist $K$-,$R$-, and $M$-computable curves which have non-computable lengths.  For $N$-computable curves, we can prove the same property by a direct construction as well. The construction needs the following simple fact.

\begin{prop}\label{prop-polygon-a<b}
Let $a< b$ and $\varepsilon$ be any positive rational numbers and let $p$ be a simple rational polygon of length $a$. There is a simple rational polygon $q$ of the length $b$ such that $d_H(p,q) \le \varepsilon$. In addition, we can choose their length-normalized parameterizations $\hat{p}$ and $\hat{q}$ such that $|\hat{p}(t) -\hat{q}(t)| \le \varepsilon$ as well.
\end{prop}
\proof
For simplicity, just consider the case $p =[(0,0), (a,0)]$. For general rational polygon $p$ we need only look at each segment of the $p$ and construct $q$ in a similar way.

Choose an integer $k>0$ such that $\max\{(b-a)/(2k),\, a/k\}  \le \varepsilon/2$. Let  $t_i= i(a/k)$ for $i\le k$ and $\varepsilon':= (b-a)/(2k)$. We define the polygon $q$ by replacing the segment $I_i:=[(t_i,0), (t_{i+1}, 0)]$ of $p$ by a polygon $q_i :=[(t_i, 0),(m_i, 0),(m_i, \varepsilon'),(t_{i+1}, \varepsilon'),(t_{i+1}, 0)]$ where $m_i=(t_i+t_{i+1})/2$. Because $l(q_i) = l(I_i) + 2\varepsilon'$, we have $l(q) = l(p) + 2\varepsilon' k = a+ (b-a)= b$. Apparently, we also have $d_H(p,q) = \varepsilon' \le \varepsilon$.

\begin{figure}[h]
\begin{center}
\begin{tikzpicture}
\draw[help lines] (-1,-1) grid (13,2);
\draw [very thick] (0,0) -- (2, 0) -- (2,1) -- (4,1) --(4,0) --(6,0)--(6,1)
--(8,1) --(8,0) --(10,0)--(10,1) --(12,1)--(12,0);

\draw[fill] (0,0) circle [radius=0.08];
\node[below] at (0,0) {$(0,0)$};

\draw[fill] (4,0) circle [radius=0.08];
\node[below] at (4,0) {$(t_{1},0)$};
\draw[fill] (6,0) circle [radius=0.08];
\node[below] at (6,0) {$(m_{1},0)$};
\draw[fill] (6,1) circle [radius=0.08];
\node[above] at (6,1) {$(m_{1},\varepsilon')$};
\draw[fill] (8,1) circle [radius=0.08];
\node[above] at (8,1) {$(t_{2},\varepsilon')$};

\draw[fill] (8,0) circle [radius=0.08];
\node[below] at (8,0) {$(t_{2},0)$};
\draw[fill] (12,0) circle [radius=0.08];
\node[below] at (12,0) {$(a,0)$};
\end{tikzpicture}
\end{center}
\caption{The polygon $q$\ ($k=3$)}\label{fig-q}
\end{figure}

\noindent Finally we look at the length-normalized parameterization $\hat{p}$ and $\hat{q}$. By construction, we have $\hat{p}(t_i) = \hat{q}(t_i)$ and $|\hat{p}(m_i)-\hat{q}(m_i)| = \varepsilon'$ for all $i$. Because the length $l(I_i) = a/k \le \varepsilon/2$ and $\varepsilon' \le \varepsilon/2$, we have $\max\{d(\hat{p}(t),\, \hat{q}(s)|\} \le \sqrt{(\varepsilon/2)^2+ (\varepsilon/2)^2}=\varepsilon$ for all $s,t \in [t_i, t_{i+1}]$. This implies immediately that $|\hat{p}(t) -\hat{q}(t)| \le \varepsilon$ for all $t\in[0,1]$.
\qed

\begin{thm}\label{thm-N-noncomp-length}
For any left computable real number $l$, there is an $N$-computable curve with the length $l$.
\end{thm}
\proof Let $l$ be a left computable real number and let $(l_n)$ be an increasing computable sequence of rational numbers which converges to $l$. W.l.o.g., we assume that $l$ and $l_n$ are positive. By Proposition \ref{prop-polygon-a<b}, we can construct a computable sequence $(p_n)$ of rational polygons with $l(p_n)=l_n$ and  $d_H(p_n, p_{n+1}) \le 2^{-n}$ inductively as follows.

First, let $p_0 = [(0,0), (l_0, 0)]$. For any $n$, if $p_n$ is already defined with $l(p_n) =l_n$, then define a new polygon $p_{n+1}$ according to the construction of Proposition \ref{prop-polygon-a<b} such that $l(p_{n+1}) =l_{n+1}$, $d_H(p_n, p_{n+1}) \le 2^{-n}$ and $|\hat{p}_n(t) -\hat{p}_{n+1}(t)| \le 2^{-n}$ where $\hat{p}_n$ is a length-normalized parameterization of $p_n$. This implies that the limit $f(t) =\lim \hat{p}_n(t)$ is a computable function which is a length-normalized parameterization of the limiting curve $p:= \lim p_n$.

Furthermore, when we construct the polygon $p_{n+1}$ from $p_n$, we should choose the constant $\varepsilon'$ (of the proof of Proposition \ref{prop-polygon-a<b}) to be smaller than one fourth of all line segments of $p_n$. In addition we should also choose the extension direction of $p_{n+1}$ carefully. In this way, we can prove by induction that, there is a constant $c$ such that $|\hat{p}(t_1)- \hat{p}_n(t_2)| \ge c\cdot |t_1-t_2|$ for all $t_1, t_2 \in [0,1]$. This concludes that $\hat{p}$ is an injective function and hence $p$ is $N$-computable.
\qed

In fact, many curves we are familiar with in mathematics have computable length. The following lemma gives a simple sufficient condition that a curve has computable length.

\begin{lem}\label{lem-complength-diff-function}
If an injective parametrization of a simple curve $C$ has a computable derivative, then $C$ has computable length.
\end{lem}
\proof Let $f(t) :=\left< x(t), y(t)\right>$ be a one-to-one parametrization of $C$ such that the derivative $f'(t)=\left< x'(t), y'(t)\right>$ is computable as well. Then the arc length of $C$ can be calculated by $l(C) = \int_0^1 \sqrt{(x'(t))^2 +(y'(t))^2} dt$ which is computable and $g(t):=\int_0^t \sqrt{(x'(x))^2 +(y'(x))^2} dx$  is a computable length-normalized parametrization of $C$.
\qed

Thus, by Lemma \ref{lem-complength-diff-function},  line segments connecting two computable points, computable polygons (connecting finitely many computable points by straight lines), computable circles, etc, all have computable length.


\section{Point Separable Classes of Curves}\label{sec-point-separable}

The main goal of this paper is to distinguish different versions of computable curves introduced in Section \ref{sec-comp-curve} in a very strong sense, i.e., by means of point separability. In this section we will introduce formally the notion of point-separability and explore some basic facts about it. Finally we show a technical lemma which are useful in the proofs of our point-separability results.

The non-equivalence of the $R$-computability and the $M$-computability of curves is proved by Gu, Lutz and Mayordomo in \cite{GLM11}. Actually they have shown that there is a polynomial time computable curve $\bf \Gamma$  which does not have any injective computable parametrization. In other words, any computable parametrization $f$ of the curve $\bf \Gamma$ must be retraced in the sense that $f(I_1) = f(I_2)$ for some disjoint subintervals $I_1, I_2 \subseteq [0; 1]$. Thus, the curve $\bf \Gamma$ is $R$-computable but not $M$-computable. In the same paper, Gu, Lutz and Mayordomo asked  whether  there exists a point which lies on a computable curve of finite length but not on any computable curve of computable length?  This leads naturally to the following notion.

\begin{defi}\label{def-point-sep}
Let $\IC$ and $\IB$ be classes of curves.
\begin{enumerate}[(1)]
    \item A point $x$ is called  {\em $\IC$-reachable} if $x$ lies on some curve $C$ of the class $\IC$.
    \item The class $\IC$ is called {\em point-separable} from the class $\IB$ if there is a $\IC$-reachable point which is not $\IB$-reachable.
\end{enumerate}
\end{defi}

\noindent Thus, if $\IC_F$ and $\IC_C$ are the classes of computable curves of finite and computable length, respectively, then, the question of Gu, Lutz and Mayordomo becomes whether $\IC_F$ is point-separable from $\IC_C$.

Notice that the endpoints of a computable curve are computable, so we can always extend a computable curve from one end so that it starts from the origin. Thus, for  $A \in \{K, R, M, N\}$, the $\IC_A$-reachable points are just those points on the plane which can be accessed from the origin along some $A$-computable curve.

If $\IC$ is the class of all planar curves, then all points are $\IC$-reachable. For some special classes of curves we can prove the point-separability very easily. For example, let $\IC$ be the class of all rational circles (i.e., centered at rational points with rational radii) and let $\IB$ be the class of all rational polygons. Then $\IC$ is point separable from $\IB$ and vice versa. The proof is quite simple. Given a rational circle $C$, any rational line segment intersects the circle $C$ in at most two points. The number of rational line segments is countable. Since the circle $C$ contains uncountably many points, there must be points on $C$ which do not lie on any rational line segment. Therefore, $\IC$ is point separable from $\IB$. The other direction can be proved similarly.

This example can be easily extended to the following proposition.

\begin{prop}\label{prop-p-sep-simple-case}
Let $\IC$ and $\IB$ be countable classes of curves such that for any curve $C\in\IC$ and $B \in \IB$, $C$ intersects $B$ at most in countably many points. Then $\IC$ is point-separable from $\IB$.
\end{prop}

It makes more sense if $\IC$ is point-separable from some subclass $\IB \subseteq \IC$. In this case, $\IC$ contains some curve which is significantly more complicated than any curve of $\IB$. To prove such kind of point-separability, the following technical lemma is very useful. It is based on a simple observation that, if a curve $C$ is not contained in another curve $C'$, then there must be a small neighborhood of some point on $C$ which is disjoint from $C'$.

\begin{lem}\label{lem-two-curves}
Let $C$ and $C'$ be two rectifiable, simple curves and let $g:[0;1]\rightarrow\IR^2$ be a parametrization of $C'$. If we have $C'\cap V_\varepsilon (z) \neq \emptyset$ for all points $z\in C$ and all open neighborhoods $V_\varepsilon (z)$, then there exists an interval $[a;b]\subseteq [0;1]$ such that $g([a;b])=C$.
\end{lem}
\proof
Suppose that $C, C'$ are rectifiable, simple curves. If $C'\cap V_\varepsilon(z) \neq \emptyset$ for any point $z\in C$ and any $\varepsilon>0$, then $C$ must be a part of $C'$, i.e., $C \subseteq C'$. Otherwise, by the compactness of $C'$, we can find a point $z$ in $C\backslash C'$ which has positive distance from $C'$ and hence some open neighborhood of $z$ is disjointed from $C$ which contradicts the hypothesis.

As a rectifiable simple curve $C'$ has an injective parametrization $f:[0;1] \to C'$. Its inverse function $f^{-1}$ is also continuous which maps particularly two end points of $C$ to $u, v\in [0;1]$. Suppose w.l.o.g. that $u<v$. Then we have $f([u;v]) =C$ due to the connectedness of the curve.

Let $h:[0;1] \to [0;1]$ be the continuous function defined by $h:= f^{-1} \circ g$. Since $f([u;v]) =C \subseteq C' =g([0;1])$, we have $[u;v] \subseteq h([0;1])$. By the continuity of $h$, there exist  $a \in h^{-1}(u)$ and $b \in h^{-1}(v)$ such that $h([a; b]) =[u; v]$ (we suppose w.l.o.g that $a < b$). This implies immediately that $g([a; b]) =C$.
\qed

By Lemma \ref{lem-two-curves}, if a curve $C$ is not contained in another curve $C'$, then there exist a point $z$ of $C$ and a neighborhood $V_\varepsilon(z)$ which is disjoint from the curve $C'$. Particularly, if $C$ is longer than $C'$, then $C$ cannot be contained in $C'$. If in addition $C$ is a rational polygon and $C'$ is a computable curve, then the point $z$ and the number $\varepsilon$ can be even rational. Thus, just by ``checking and waiting'' we can always find effectively such a rational point $z$ and the corresponding rational neighborhood $V_\varepsilon(z)$. This idea will be used several times in the proofs of Section \ref{sec-main}. In those proofs, we are given a rational polygon $C$ and a ($K$-, $R$-, $M$- or $N$-)computable curve $C'$. As long as we can verify that $C$ is sufficiently different from $C'$ (and hence $C$ is not contained in $C'$), then we can always find a point $z$ on $C$ and an $\varepsilon$-neighborhood $V_\varepsilon(z)$ which is disjoint from $C'$.

\section{Length-Normalized Computable Curves}\label{sec-n-comp-curve}

An $N$-computable curve has a length-normalized computable parametrization. This type of computable curves model the particle motion of constant speed. By Theorem \ref{thm-N-noncomp-length}, an $N$-computable curve does not necessarily have a computable length. Thus, the class $\IC_N$ is a proper superset of $\IC_C$. Our next result shows that the class $\IC_N$ is different from the class $\IC_C$ in a very strong way. Namely, for any curve $C \in \IC_N$, $C$ is either an element of $\IC_C$, or any non-trivial segment of $C$ is not in $\IC_C$.

\begin{lem}\label{lem-N-C-no-containing}
If $C$ is an $N$-computable curve of non-computable length, then no non-trivial segment of $C$ is a computable curve of computable length.
\end{lem}
\proof
Let $C$ be an $N$-computable curve of length $l$ which is not computable. By Definition \ref{Def-comp-curve}, there is an injective computable function $f:[0,1] \to \IR^2$ such that $\rang(f) =C$ and $l(f[0,t]) = t\cdot l$ for all $t \in [0,1]$.

If $C_1\subseteq C$ is a nontrivial segment of $C$, then there are $t_1 < t_2$ in $[0,1]$ such that $f[t_1, t_2] =C_1$. Suppose by contradiction that $C_1$ is a computable curve of computable length $l_1$. Then it must be also $N$-computable by Theorem \ref{thm-K+compLengh}, and it has a normalized computable parametrization $f_1:[0,1]\to C_1$. Let $A:=f_1(0)$ and $B:=f_1(1)$ be the endpoints of $C_1$. Both $A$ and $B$ are computable points. Because $f$ is an injective computable function and $A=f(t_1)$ and $B=f(t_2)$, the numbers $t_1, t_2 \in [0,1]$ are also computable.

Let $l_0$ be the length of the segment $f[0, t_1]$. Then, we have $l_0= t_1\cdot l$ and $(l_0 +l_1) = t_2 \cdot l$. This implies that $l_1 =(t_2-t_1)l$. Therefore $l = l_1/(t_2-t_1)$ is computable which contradicts the hypothesis.
\qed

From a mathematical point of view, Lemma \ref{lem-N-C-no-containing} is quite surprising and even strange. Physically, an $N$-computable curve $C$ can model the algorithmic particle motion of a constant speed. In particular, if the argument $t$ of its parametrization $f:[0,1]\to C$ is regarded as the time, the length $l$ corresponds to the speed of the motion. Thus, an $N$-computable curve of non-computable length is a model of a particle motion with non-computable constant speed, while its trace can be effectively determined. In this case, of course, any of its segments models also a particle motion of a non-computable constant speed.

From Lemma \ref{lem-N-C-no-containing} we can prove the following point-separable result.

\begin{thm}\label{thm-N-C-sep}
There is an $N$-computable curve $K$ and a point on $K$ which is not on any computable curve of computable length. That is, the classes $\IC_N$ and $\IC_C$ are point-separable.
\end{thm}

\proof
Let $K$ be an $N$-computable curve of a non-computable length and let $\IC_C =\{C_i: i\in\IN\}$ be a (not necessarily effective) enumeration of all computable curves of computable length. By Lemma \ref{lem-N-C-no-containing}, the intersection $B_i := K \cap C_i$ is a nowhere dense set for any $i$. Thus, the set $B := \bigcup_{i\in\IN} (K \cap C_i)$ is a meager set. This implies immediately that $K \setminus B \neq \emptyset$. That is, there is a point on $K$ which is not on $C_i$ for all $i\in\IN$.
\qed

Theorem \ref{thm-N-C-sep} answers, even in a stronger sense, the question of Gu, Lutz and Mayordomo \cite{GLM11} that wether there exists a point which lies on a computable curve of finite length but is not covered by any computable curve of computable length, because their notion of computable curves
is the $R$-computable curves.


\section{Point-Separable Classes of Computable Curves}\label{sec-main}

In this section we will prove the point-separability of four versions of the computable curves. The proofs are standard finite injury priority constructions. We sketch only the main ideas, because a priority construction with complete formal details, although it is technically not difficult, will be very long and could hide the essential proof ideas. The detailed explanation about the injury priority construction can be found in \cite{Soa87}.

Remember that a function $f:[0,1] \to \IR^2$ is computable if there is a Turing machine $M$ which computes $f$ in the sense that $M$ transfers any sequence $(t_s)$ of rational numbers which converges effectively to $t\in [0,1]$ to a sequence $(z_s)$ of rational points which converges effectively to $f(t)$. Equivalently, $f$ is computable if and only if there is a computable sequence $(p_n)$ of rational polygon functions $p_n:[0,1] \to \IR^2$ which converges to $f$ uniformly and effectively. For technical simplicity, we can understand in this section that a Turing machine $M$ computes a function $f$ means that $M$ computes a sequence $(f_s)$ of rational polygon functions which converges to $f$ uniformly effectively, i.e., $|f(t) -f_s(t)| \le 2^{-s}$ for all $s\in\IN$ and $t\in [0,1]$.

Let $(M_e)$ be an effective enumeration of all Turing machines such that $M_e$ possibly computes a computable sequence $(\varphi_{e,s})_s$ of rational polygon functions defined on $[0,1]$ in the sense that $M_e(s) = \varphi_{e,s}$. If the sequence $(\varphi_{e,s})_s$ converges to $\varphi_e$ uniformly effectively, then $M_e$ computes the function $\varphi_e$ which can be regarded as a parameterization of an $R$-computable curve $C_e$. The polygon curve defined by $\varphi_{e,s}$ is denoted by $C_{e,s}$. If $M_e$ doesn't compute a computable sequence of rational polygons, or the sequence doesn't converge effectively, then we say that $M_e$  computes only an empty curve, i.e., $C_e =\emptyset$. Therefore, $(\varphi_e)$ is an effective enumeration of all $R$-computable curves.

Now we are ready to show that the classes $\IC_K$ and $\IC_R$ are point-separable. Our proof will use the following fact about the ``sweep" of a continuous function.

Let $f:[a,b]\to \IR^2$ be a continuous function, let $q\in \rang(f)$ be a point and $\delta >0$ be a constant. An interval $[t_0, t_3]\subseteq [a,b]$ is called a $(q, \delta)$-sweep of $f$ if there is a point $p$ in the range of $f$ such that  $|p-q| =\delta$ and the function $f$ travels from $q$ to $p$, turns back to $q$ and then go through $p$ and forward again. In other words, there are $t_1, t_2$ with $t_0<t_1<t_2<t_3$  such that $f(t_0) =f(t_2) =q$, $f(t_1) =f(t_3)=p$ and $f[t_0,t_1] =f[t_1, t_2], = f[t_2, t_3]$.  In other words, $f$ retraces the curve segment between $q$ and $p$ two times.

\begin{lem}\label{lem-sweep}
Let $a<b$ and let $f:[a,b] \to \IR^2$ be a continuous function. For any constant $\varepsilon>0$, there can be at most finitely many $(q,\delta)$-sweeps where $q\in \rang(f)$ and $\delta\ge \varepsilon$.
\end{lem}

\proof
Since $f$ is also uniformly continuous on the interval $[a, b]$, there exists a $\delta' >0$ such that $|f(t)-f(t')| < \varepsilon$ if $|t-t'|< \delta'$. Now, suppose that $[t_0, t_3]\subseteq [a,b]$  is a $(q, \delta)$-sweep of $f$ for some $q\in \rang(f)$ and  $\delta \ge \varepsilon$, then we have $|t_0 -t_3| \ge 3\delta'$. If there is another $(q, \delta)$-sweep $[s_0, s_3]$ which is, say, inside the interval $[t_0, t_1]$, then if forces the interval length $|t_3-t_0|$ to be greater than $5\delta'$. Therefore, any $(q, \delta)$-sweep, no matter nested or not,  costs at least a length $2\delta'$ of the interval $[a,b]$.  This implies immediately that the finite interval $[a,b]$ can contain only finitely many such sweeps.
\qed

\begin{thm}\label{Thm-K-R-sep}
There exists a rectifiable $K$-computable curve $K$ and a point $z$ on $K$ such that $z$ does not belong to any $R$-computable curve $C$.
\end{thm}
\proof By Corollary \ref{cor-K-equivalence}, a rectifiable curve $K$ is $K$-computable iff there is a computable sequence $(K_n)$ of rational polygons which converges to $K$ effectively in the sense that $d_H(K_n, K) \le 2^{-n}$ for all $n$. In the following, we will construct such a computable sequence $(K_n)$ of rational polygons which converges effectively to a curve $K$, and at the same time we construct also a computable sequence $(z_n)$ of rational points which converges to a point $z$ on $K$. Let $K_s$ and $z_s$ be the candidates  constructed at the stage $s$.

Let $(M_e)$ be an effective enumeration of all Turing machines and let $(C_e)$ be the corresponding enumeration of all $R$-computable curves. Thus, it suffices to guarantee that the constructed $K$-computable curve $K$ and the point $z$ on $K$ satisfy, for all $e\in\IN$, the following requirements:
\begin{eqnarray*}
  R_e &:& \mbox{The point $z$ does not belong to $C_e$}.
 \end{eqnarray*}

We explain the strategy to satisfy a single requirement $R_e$ first.

Suppose, at stage $s+1$, that a rational polygon $K_s$ and a point $z_s$ on $K_s$ are defined. In addition, we have also defined a neighborhood $B_{e-1,s}$ which contains the point $z_s$ as well as part of $K_s$. The new rational polygon $K_{s+1}$ will be defined by, if it is necessary, changing part of polygon of $K_s$ within the neighborhood $B_{e-1,s}$. Meanwhile, we construct a new neighborhood $B_{e,s+1} \subseteq B_{e-1,s}$ which contains the new point candidate $z_{s+1}$ on the polygon $K_{s+1}$ such that $B_{e,s+1}$ is disjoint from the curve $C_e$. In this way, we can guarantee that the point $z:=\lim z_s$ is on the curve $K:=\lim K_s$, but not on the curve $C_e$. That is, the requirement $R_e$ is satisfied.

For simplicity, let $B_{e-1,s}$ be the box of a side-length $\delta_{e,s}\le 2^{-(2e+2)}$ centered at the point $(\delta_{e,s}/2, 0)$ and the polygon $K_s$ contained in this box is simply the line segment $J:=[(0,0),(\delta_{e,s},0)]$.  Suppose now that $\varphi_e$ is a total function and $C_e$ intersects with $B_{e-1,s}$, otherwise, we need do nothing. Let $J_e$ be the part of $C_e$ in the box $B_{e-1,s}$. If $J\not\subseteq J_e$, then, by Lemma \ref{lem-two-curves}, we can find a new neighborhood $B_{e,s+1} \subseteq B_{e-1, s}$ and  a point $z_s \in B_{e,s+1} \cap K_s$ such that $J_e\cap B_{e,s+1} = \emptyset$ and hence the requirement $R_e$ is satisfied. Notice that, $J\not\subseteq J_e$ can be determined, say,  by finding a rational point $q \in J$ such that $d_H(q, J_{e,t}) > 2^{-(t+1)}$, for some $t\le s$, where $J_{e,t}$ is the intersection of $C_{e,t}$ with $B_{e-1,s}$. Note that, we can always compute the curve $C_{e,t}$ by the computation of $M_e(t)$ ($=\varphi_{e,t}$) up to $s$ steps which is denoted by $M_{e,s}(t)$.

We consider now the case that $J\subseteq J_e$. By Lemma \ref{lem-sweep}, $\varphi_e$ can have at most finitely many $(q, \delta)$-sweeps for any $q\in J$ and $\delta \ge \varepsilon = \delta_{e,s}/4$. Therefore, there must be a rational point $q$ and an $\varepsilon_1$-neighborhood $V_{\varepsilon_1}(q)$ of $q$ such that $\varphi_e$ does not have a $(q_1, \delta)$ sweep for all $q_1 \in V_{\varepsilon_1}(q)$ and $\delta\ge \varepsilon$. We can find such a $q$ by calculating $\varphi_e$ to sufficient precision, that is, by calculating $\varphi_{e,t}$ for sufficiently large $t\le s$. Otherwise, either $\varphi_e$ is not a total function, or $J$ is not contained in $C_e$. Here, ``calculating $\varphi_e$ to sufficient precision" means we try to find a maximum $t\le s$ such that the computations $M_{e,s}(0), M_{e,s}(1), \cdots, M_{e,s}(t)$ all halt, and $t$ is large enough such that the precision $2^{-t}$ is good enough to determine the ``no-sweep" case.

Suppose that we already find the rational point $q :=(q, 0)$ such that $\varphi_e$ does not have any $(q,\varepsilon)$-sweep. Then we define the new polygon $K_{s+1}$ by replacing the linear segment $J':=[(q+\varepsilon,0),(q+2\varepsilon, 0)]$ by the polygon  $J'':=[(q+\varepsilon, 0), (q, \delta),(q+2\varepsilon, 0)]$, where $\delta:= \min\{ 2^{-(s+1)}, \varepsilon\}$. Apparently, we have $d_H(K_{s+1}, K_s) \le \delta < 2^{-s}$.

\begin{figure}[h]
\begin{center}
\begin{tikzpicture}
\draw[help lines] (-1,-1) grid (10,2);
\draw [very thick] (-1,0) -- (4, 0) -- (0,1) -- (8,0) --(10,0);

\draw[fill] (0,0) circle [radius=0.08];
\node[below] at (0,0) {$(q,0)$};
\draw[fill] (0,1) circle [radius=0.08];
\node[above] at (0,1) {$(q,\delta)$};
\draw[fill] (4,0) circle [radius=0.08];
\node[below] at (4,0) {$(q+\varepsilon,0)$};
\draw[fill] (8,0) circle [radius=0.08];
\node[below] at (8,0) {$(q+2\varepsilon,0)$};

\end{tikzpicture}
\end{center}
\caption{The polygon $J''$ which simulates a $(q,\varepsilon)$-sweep}\label{fig-K-R}
\end{figure}

After this change, the constructed new polygon $K_{s+1}$ is different enough from the curve $C_e$ so that, by Lemma \ref{lem-two-curves}, we can find a new neighborhood $B_{e,s+1} \subseteq B_{e-1, s}$ and  a point $z_{s+1} \in B_{e,s+1} \cap K_{s+1}$ such that $C_e\cap B_{e,s+1} = \emptyset$ and hence the requirement $R_e$ is satisfied. For technical reasons, we should also choose the new neighborhood $B_{e,s+1}$ small enough such that it doesn't contain any other $z_t$ for $t \le s$.

\noindent Notice that, in the above construction, the line segment $J'$ of length $\varepsilon$ of $K_s$ is replaced by $J''$ which is a polygon of two line segments of the lengths $\sqrt{\varepsilon^2 + \delta^2}$ and $\sqrt{(2\varepsilon)^2 + \delta^2}$, respectively. Therefore, the length-increment of the new polygon can be estimated as follows:
\begin{eqnarray*}
\left| l(K_{s+1}) - l(K_s)\right| &=& \sqrt{(2\varepsilon)^2 + \delta^2}+\sqrt{\varepsilon^2 + \delta^2} -\varepsilon
\le  \sqrt{(2\varepsilon)^2 + \varepsilon^2}+\sqrt{\varepsilon^2 + \varepsilon^2} -\varepsilon \\
&\le& 4\varepsilon \le 4\cdot 2^{-(2e+2)} = 2^{-2e}.
\end{eqnarray*}

To satisfy all requirements $R_e$ simultaneously, we need the technique of the finite injury priority construction. We say that a requirement $R_i$ has a higher priority than $R_j$ if $i <j$. At any stage $s$, we have to construct a finite sequence $(B_{i,s})_{i \le a_s}$ of the neighborhoods and a finite sequence of rational points $(z_{i,s})_{i\le a_s}$ for some natural number $a_s$, in addition to the rational polygon $K_s$,  such that
\begin{eqnarray*}
B_{0,s} \supseteq B_{1,s} \supseteq \cdots \supseteq B_{a_s, s} \mbox{ and } (\forall i \le a_s)( z_{i,s} \in B_{i,s} \cap K_s)
\end{eqnarray*}
and that $B_{i,s}$ is disjoint from the curve $C_i$. The neighborhood $B_{j,s}$ has to be canceled (by the fact that $a_t < j$) at some stage only if a new neighborhood $B_{i,t}$ is redefined at the stage $t$ for some $i<j$. In this case, The requirement $R_j$ is injured.  Whenever a box $B_{i,s}$ is defined according to the strategy mentioned above, it is not necessary to redefine it again unless $R_i$ is injured by a requirement of higher priority. By an simple induction it is not difficult to prove that any requirement $R_i$ can be injured no more than $2^i-1$ times and $B_i$ needs to be redefined at most $2^i$ times. Thus, $B_i := \lim_{s\to\infty} B_{i,s}$ exists and $B_i$ is disjointed from $C_i$. Similarly, $z_i := \lim_{s\to\infty} z_{i,s}$ exists too and $z_i \in B_i$.  Because $B_{i+1} \subseteq B_i$ and the size of $B_i$ converges to zero if the index $i$ goes to infinity, the limit $z:=\lim_{i\to\infty} z_i$ exists and $z$ is a point on the curve $K:= \lim_{s\to\infty} K_s$. Here the existence of the limit and the $K$-computability of the limiting curve $K$ follows from the Corollary \ref{cor-K-equivalence}. The point $z$ belongs to all neighborhoods $B_i$ and hence is disjointed from all curves $C_i$. Therefore, $z$ is a point on a $K$-computable curve but is never covered by an $R$-computable curve.

Finally, we can show that the limiting curve $K$ has a finite length. Notice that, for each $i$, the curve length can be increased by the actions for $R_i$ at most $2^i$ times, while it can increase at most $2^{-(2i)}$ each times. This means that the total length-increment caused by $R_i$ is bounded by $2^{-i}$. Therefore the total length of $K$ must be finite.
\qed

In the following, we will show that the classes of $R$-computable curves, $M$-computable curves and $N$-computable curves are all point-separable. Because the proofs are finite injury priority constructions similar to that of Theorem \ref{Thm-K-R-sep}, we just give sketches of the proofs.

\begin{thm}\label{thm-R-M-sep}
There exists a rectifiable $R$-computable curve $K$ and a point $z$ on $K$ such that $z$ does not belong to any $M$-computable curve $C'$.
\end{thm}

\proof (Sketch) We need only to construct an $R$-computable curve $K$ and a point $z$ on $K$ which satisfy, for all $i\in\IN$, the requirements
\begin{eqnarray*}
  R_i &:& \mbox{If $\varphi_i$ is an injective parametrization of $C_i$, then $z$ is not on $C_i$.}
\end{eqnarray*}
where $(\varphi_i)$ is a computable enumeration of all (possibly partial) computable functions $\varphi_i: [0,1] \to \IR^2$. The $R$-computable curve $K$ is defined as the limit of a computable sequence $(K_s)$ of rational polygons which converges to $K$ effectively in the sense that $d_H(K, K_s) \le 2^{-s}$ for all $s$. At the same time, we also construct a computable sequence $(f_s)$ of real functions $f_s:[0,1]\to \IR^2$ such that $f_s$ is a computable parametrization of $K_s$, and the sequence $(f_s)$ converges effectively to a computable function $f$ which is a parametrization of $K$. This guarantees that $K$ is an $R$-computable curve. In addition, we construct a sequence $(z_s)$ of points such that $z_s$ is on the polygon $K_s$ and disjoint from $C_i$, and $(z_s)$ converges to a point $z$ on $K$.

The sequences $(K_s)$, $(f_s)$ and $(z_s)$ are constructed in stages by a finite injury priority method. We explain the idea of how to satisfy a single requirement $R_i$ only.

Suppose that, at some stage $s$, we have defined a rectangular box $B_i$ of a side length $a:=2^{-b}$ which contains a segment of the polygon $K_s$ constructed so far, where $b = \max\{s, 2i+3\} $. For simplicity, let $B_i$ be the box centered at the point $(a/2, 0)$ and let $l_0= [(0,0),(a, 0)]$ be the line segment of the polygon $K_s$ in $B_i$. Suppose also that the parametrization $f_s$ defined at the stage $s$ retraces the line $l_0$ three times, that is, it starts at $(0,0)$, goes to $(a, 0)$, back to $(0,0)$ and then goes forward to $(a, 0)$. Because the length of $l_0$ is bounded by $2^{-s}$, we can always define $f_s$ in this way without violating the effective convergency of the sequence $(f_s)$.

Similar to the proof of Theorem \ref{Thm-K-R-sep}, calculating $\varphi_i$ to sufficient precision so that we can determine the following cases.

Case 1. If $C_i$ is disjoint from $B_i$, then we need to do nothing.

Case 2. If $C_i$ intersects the box $B_i$ and $C_i$ closely passes the segment $l_0$ only once. In this case, replace the segment $l_0$ by a Z-sweep $l_1$ of height $\delta$: $l_1 :=[(0,0),(a, \delta),(0, -\delta),(a, 0)]$. Where $\delta>0$ is a sufficiently small rational number.

\begin{figure}[h]
\begin{center}
\begin{tikzpicture}
\draw[help lines] (-1,-2) grid (9,2);
\draw [very thick] (-1,0) -- (0, 0)-- (8,1) --(0,-1)--(8,0) --(9,0);

\draw[fill] (0,0) circle [radius=0.08];
\node[above] at (0,0) {$(q,0)$};
\draw[fill] (8,1) circle [radius=0.08];
\node[above] at (8,1) {$(a,\delta)$};

\draw[fill] (0,-1) circle [radius=0.08];
\node[below] at (0,-1) {$(0,-\delta)$};
\draw[fill] (8,0) circle [radius=0.08];
\node[below] at (8,0) {$(a,0)$};

\end{tikzpicture}
\end{center}
\caption{A Z-sweep polygon $l_{1}$}\label{fig-R-M}
\end{figure}

Case 3. $C_i$ is close to $l_0$ and also has Z-sweep near $l_0$. Notice that $l_0$ dose not have a Z-sweep. We do nothing in this case.

In both cases 2 and 3, since the new polygon $l_1$ is sufficiently different from $C_i$, by Lemma \ref{lem-two-curves}, we can choose a new box $B'_i \subseteq B_i$ which contains part of $l_1$ and choose a point $z$ on $l_1$ in the box $B'_i$. This new box $B'_i$ and the new point $z$ can be used as witnesses for the requirement $R_i$. In addition, to guarantee the finite length of the limiting curve, we should choose $\delta \le \min\{a, 2^{-(s+1)}\}$ if it is implemented at the stage $s$. Then we can redefine the parametrization $f_{s+1}$ of $K_{s+1}$ (which contains $l_1$) such that  $|f_{s}(t) -f_{s+1}(t)| \le 2^{-{(s+1)}}$. This is possible because the original parametrization $f_s$ traces the $l_0$ three times which is very close to the Z-sweep of $l_1$. This guarantees that the function sequence $(f_s)$ converges effectively.

On the other hand, we can estimate the length of $l_1$ as follows:
\begin{eqnarray*}
l(l_1) &=& 2\sqrt{a^2 + \delta^2} + \sqrt{a^2 + (2\delta)^2} \le (2\sqrt{2} +\sqrt{5})a \le 6a
\end{eqnarray*}
Therefore, the length of the polygon $l_1$ differs from $l_0$ by no more than $5a \le 2^{-(2i)}$. Since the requirement $R_i$ will be injured at most $2^i-1$ times and the curve can be increased due to the strategy for $R_i$ at most $2^i$. Thus, the limiting curve is of a finite length.

The strategy described above can be used to satisfy all requirements $R_i$ simultaneously by a finite injury priority method. The detailed construction is very similar to the proof of Theorem \ref{Thm-K-R-sep} and is omitted here.
\qed

Finally, we show the difference between $M$- and $N$-computability of curves.

\begin{thm}\label{thm-M-N-sep}
There exists a rectifiable  $M$-computable curve $K$ and a point $z$ on $K$ such that $z$ does not belong to any $N$-computable curve $C'$. That is, the classes $\IC_M$ and $\IC_N$ are point-separable.
\end{thm}
\proof (Sketch) We will use the priority technique again to construct an $M$-computable curve $K$ and a point $z$ on $K$ such that the following requirements are satisfied
\begin{eqnarray*}
  R_i &:& \mbox{If $\varphi_i$ is a length-normalized parametrization of $C_i$, then $z$ is not on $C_i$.}
\end{eqnarray*}
Again, we want to construct a computable sequence $(K_s)$ of rational polygons and a computable sequence $(f_s)$ of injective functions which converges to the curve $K$ and the computable function $f$, respectively, such that $f$ is an injective parametrization of $K$. At the same time, we also construct a sequence $(z_s)$ of points which converges to a point $z$ on $K$, but $z$ is disjoint from all $C_i$.

The strategy for satisfying a single requirement $R_i$ is to find a neighborhood $B_i$ which is disjoint from $C_i$ and which contains a segment of $K$ and a point $z$ on this segment. To guarantee that the curve $K$ has a finite length, similar to the proofs of Theorem \ref{Thm-K-R-sep} and \ref{thm-R-M-sep}, we should choose $B_i$ so that the length of the curve is increased at most  $2^{-(2i)}$. For simplicity, suppose that $B_i$ is a neighborhood of size $2^{-2i}$ centered at $(2^{-(2i+1)},0)$ and let $B_i$ be our first candidate of the witness neighborhood. Suppose in addition that the line segment $J$ connecting $(0,0)$ and $(2^{-2i}, 0)$ is the segment of $K$ in the box $B_i$. Let $f'$ be an injective computable parametrization of the (current candidate of) $K$.

Suppose that $C_i$ is an $N$-computable curve and $\varphi_i$ is a length-normalized parametrization of $C_i$. Just wait until the Turing machine $M_i$ can compute  $\varphi_i$ to sufficient precision. As long as $C_i$ is disjoint to $B_i$, we need to do nothing. If $C_i$ does intersect with $B_i$, but is not close to the segment $J$, then we can apply Lemma \ref{lem-two-curves} to choose a new neighborhood $B'_i \subseteq B_i$ which contains part of $J$ but is disjoint from $C_i$. Otherwise, suppose that $C_i$ is very close to the segment $J$. That is, there are $t_1, t_2\in [0,1]$ such that the segment $\varphi_i([t_1, t_2])$ almost coincides with $J$. Then compute the middle point $\varphi_i((t_1+t_2)/2)$ of the segment $\varphi_i([t_1, t_2])$  and check if it is close to the middle point of $J$. If it is not the case, then $\varphi_i$ is not length-normalized and we are done. Otherwise, replace the segment $J$ by a polygon $J'$ which double the length of the first half of the segment $J$ (i.e. the part from $(0,0)$ to $(2^{-(2i+1)}, 0)$) by introducing small zigzags like the graph in Figure \ref{fig-M-N}.

\begin{figure}[h]
\begin{center}
\begin{tikzpicture}
\draw[help lines] (-1,-1) grid (11,1);
\draw [very thick] (0,0) -- (1/2, 0)-- (1/2,1/2) --(1,1/2)--(1,0) --
(1.5,0)--(1.5,1/2)--(2,1/2)--
(2,0) --(2.5,0)--(2.5,1/2)--(3,1/2)--
(3,0) --(3.5,0)--(3.5,1/2)--(4,1/2)--
(4,0) --(4.5,0)--(4.5,1/2)--(5,1/2)--
(5,0) --(10,0);

\draw[fill] (0,0) circle [radius=0.08];
\node[below] at (0,0) {$(0,0)$};
\draw[fill] (5,0) circle [radius=0.08];
\node[below] at (5,-0.2) {$(2^{-(2i+1)},0)$};

\draw[fill] (10,0) circle [radius=0.08];
\node[below] at (10,-0.2) {$(2^{-2i},0)$};

\end{tikzpicture}
\end{center}
\caption{A new polygon $J'$ with doubled length of the first half segment.}\label{fig-M-N}
\end{figure}

\noindent Denote the new (whole) polygon by $K_s$. At the same time, modify the function $f'$ to a new injective function $f_s$ such that $f_s$ is a computable parametrization of $K_s$.  Now the part of $K_s$ in $B'_i$ is different enough from the curve $C_i$ and hence we can apply Lemma \ref{lem-two-curves} to find a point on $K_s$ and a neighborhood $B'_i \subseteq B_i$ of $z$  which is disjoint from $C_i$.

By a standard priority construction, the curve $K$ can be constructed as the effective limiting curve of a computable sequence $(K_s)$ of rational polygons, and $K$ has an injective computable parametrization $f$ which is the limit of a computable sequence $(f_s)$ of injective functions. In addition, the limit $z:= \lim z_s$ is a point on $K$ which is disjoint from any $C_i$, if $C_i$ is $N$-computable.
\qed

\end{document}